\documentclass[12pt]{article}
\usepackage[T2A]{fontenc}
\usepackage[cp866]{inputenc}
\usepackage{latexsym}
\usepackage {psfig}
\usepackage {times}
\newcommand\be {\begin {equation}}
\newcommand\ee {\end {equation}}
\newcommand\beq {\begin {eqnarray}}
\newcommand\eeq {\end {eqnarray}}
\newcommand\bc {\begin {center}}
\newcommand\ec {\end {center}}
\def\d {{\rm d}}

\def\disp {\displaystyle}
\def\prt {\partial}
\def\intl{\int\limits}
\def\eps {\varepsilon}
\def\B {{\rm B}}
\def\C {{\rm C}}
\def\D {{\rm D}}
\def\G {{\rm G}}
\def\vf {\mbox {\boldmath $f$}}
\def\vv {\mbox {\boldmath $v$}}
\def\vx {\mbox {\boldmath $x$}}
\def\vchi{\mbox {\boldmath $\chi$}}
\def\Dr#1#2{\frac {\prt#1}{\prt#2}}
\def\DR#1#2{\frac {\d#1}{\d#2}}
\title {\bf Zeldovich Flow on Cosmic Vacuum Background:\\
New Exact Nonlinear Analytical Solution}

\author{Arthur D.~Chernin$^{1,2,3}$, Dmitriy I. Nagirner$^4$,
Svetlana V.~Starikova$^4$ \\
$^1$Sternberg Astronomical Institute, Moscow University, Moscow, 119899,
Russia,\\
$^2$Tuorla Observatory, University of Turku, Piikki\"o, FIN-21500, Finland,\\
$^3$Astronomy Division, Oulu University, Oulu, FIN-90014, Finland,\\
$^4$Sobolev Astronomical Institute, St.-Petersburg State University,\\
Staryi Peterhoff, 198504, St.-Petersburg, Russia
}
\date{~}

\begin {document}

\maketitle  

\begin{abstract}
A new exact nonlinear Newtonian solution for a plane matter flow
superimposed on the isotropic Hubble expansion is reported. The dynamical
effect of cosmic vacuum is taken into account. The solution describes the
evolution of nonlinear perturbations via gravitational instability of matter
and the termination of the perturbation growth by anti-gravity of vacuum
at the epoch of transition from matter domination to vacuum domination.
On this basis, an `approximate' 3D solution is suggested as an analog of
the Zeldovich ansatz.
\end{abstract}

\newpage
\section {Introduction}
Three decades ago, Zeldovich (1970) published in this {\it Journal} his now
famous nonlinear theory of gravitational instability formulated in terms of
Newtonian mechanics. The theory was first used for pancake cosmogony by
Zeldovich and his group (see for a review Shandarin\& Zeldovich 1989). Later on
it was realized that the theory can be judiciously applied to a wide range of
cosmological scenaria; in particular, the dynamics given by the theory
underlies gravitational clustering (Peebles 1993). The Zeldovich theory
provides also an effective analy\-ti\-cal tool for optimization of cosmological
simulations (Melott 1993).

The Zeldovich theory is based on an exact analytical solution which
describes a plane pressure-free matter flow superimposed on the regular Hubble
expansion:
\begin{equation} \label{eq:x1x2x3}
x_1 = a(t) \chi_1 + b(t) \beta(\chi_1), \;\;\;
x_2 = a(t) \chi_2, \;\;\;
x_3 = a(t) \chi_3,
\end{equation}
where $\vx=(x_1,x_2,x_3)$ are the Eulerian coordinates
(Car\-te\-si\-an) of an
 element (`particle') with a Lagrangian coordinates
$\vchi=(\chi_1,\chi_2,\chi_3)$, and $\beta(\chi_1)$ is an arbitrary function of
only one coordinate $\chi_1$. The first term in the first equation represents
--- together with the two other equations --- the unperturbed isotropic
solution for the Hubble flow, and the second term in the first equation
represents a perturbation flow which depends on $\chi_1$ and time $t$.

The solution was obtained by Zeldovich for the parabolic expansion (a spatially
flat model), in which the scale factor
$\disp a(t)\!=\!a_0\!\left(\frac{3}{2}H_0 t\right)^{2/3}$, where $a_0$ and
$H_0$ are the present values of $a(t)$ and Hubble constant, $b(t)\!=\!t^{4/3}$.
The perturbation corresponds to the growing mode of Lifshits (1946) linear
solution for gravitational instability of non-relativistic matter and has the
same time dependence as in the linear theory.

A decreasing mode of Lifshits theory can also be introduced to
the solution (Zentsova and Chernin 1980), and then one has instead of
 the first
equation in Eq.(\ref{eq:x1x2x3}):
\begin{equation} \label{eq:x1ZCh}
x_1(\chi_1,t) = a(t) \chi_1 + b(t) \beta(\chi_1) + c(t) \gamma(\chi_1),
\end{equation}
where $c(t) = t^{-1/3}$, and $\gamma(\chi_1)$ is the second arbitrary function
of $\chi_1$. For this more general solution, which is also nonlinear and
exact, the density is
\begin{equation} \label{eq:rho}
\rho\!=\!\frac {a_0}{6\pi Gt^2}\left[a_0\!+\!
\left(\frac {3}{2}H_0\right)^{-2/3}\left(t^{2/3}\Dr {\beta}
{\chi_1}\!+\!t^{-1}
\Dr {\gamma}{\chi_1}\right)\right]^{-1}\!.
\end{equation}

In Zeldovich (1970) work, a 3D `approximation' is suggested, in which 3D
perturbation flow is assumed to have the same structure and time dependence 
as
the plane perturbation flow in the exact solution of Eq.(\ref{eq:x1x2x3}).
This ansatz has been justified by many numerical simulations (see again
Shandarin and Zeldovich 1989). The corresponding exact analytical 3D
nonlinear solution of the problem has not been found yet and there is little
hope to find it.

In this {\it Letter}, we report a generalization of Zeldovich theory by
accounting for the dynamical effect of the cosmological constant $\Lambda$, or
cosmic vacuum. We give a new exact nonlinear Newtonian solution with the same
symmetry as in Eqs.(\ref{eq:x1x2x3}--\ref{eq:rho}), but for a non-zero
cosmological constant, or vacuum density $\disp\rho_{\Lambda} = \frac{\Lambda
c^2}{8 \pi G}$. The observational basis for this approach is due to the recent
studies of distant type Ia supernovae (Perlmutter et al. 1998, Riess et al.
1999) which indicate that the vacuum density (in the units of the critical
density) is $\Omega_{\Lambda} = 0.7 \pm 0.1$, which is in concordance with the
bulk of observational data that come also from the cosmic age, the cosmic
microwave background anisotropy in combination with cluster dynamics, etc.
(Carrol et al. 2000, Wang et al. 2000).

In spherical symmetry, the nonlinear solution of the same problem (without
vacuum) is given in the frame of the well-known Tolman-Bondi-Lema\^\i tre
model (see Lahav et al. 1991 and the references therein).

\section {Basic Equations}

Following Zeldovich (1970), we consider a plane perturbation flow 
of pressure-free matter and search for a solution in the Lagrangian form

\begin{equation} \label{eq:x1x2x3new}
x_1 = a(t) \chi_1 + \delta(\chi_1,t), \;\;\;
x_2 = a(t) \chi_2, \;\;\;
x_3 = a(t) \chi_3.
\end{equation}

The equation of motion for the `unperturbed' function $a(t)$ in the background
Friedman solution contains the matter and vacuum densities
\begin{equation} \label{eq:ddota}
\ddot {a} = -\frac {4\pi G}{3}\rho_\G a,
\end{equation}
where the effective gravitating density
\begin{equation} \label{eq:roGdef}
\rho_\G =\rho_0\frac {a_0^3}{a^3} + \rho_{\Lambda} + 3\frac {P_{\Lambda}}{c^2}=
\rho_0\frac {a_0^3}{a^3} - 2 \rho_{\Lambda}.
\end{equation}
Here $\rho_0$ is the present-day mean matter density,
$P_{\Lambda}$ is the va\-cu\-um pressure, and the equation of
state of vacuum is $P_{\Lambda} = - \rho_{\Lambda} c^2$.

The first integral of Eq.(\ref{eq:ddota}),
\begin{equation} \label{eq:dota2}
\dot {a}^2 = \frac {8\pi G}{3}\left(\rho_0\frac {a_0^3}{a^3}+
\rho_{\Lambda}\right)a^2 - kc^2,
\end{equation}
is the Friedman cosmology equation, where $k = 1, 0, -1$ for
elliptic,parabolic and hyperbolic dynamics, respectively.

The perturbation plane flow is considered in the Newtonian approximation.
The basic equations for the flow of Eq.(4) are the Euler equation of motion,
the continuity equation and the Poisson equation:
\beq \label{eq:eqmove}
& \strut\disp \dot {\vv}+ (\vv\nabla)\vv = -\nabla\varphi, & \\
 \label{eq:eqcont}
& \strut\disp \dot {\rho} + \nabla(\rho\vv) = 0, & \\ \label{eq:Poisson}
& \strut\disp \triangle\varphi = 4\pi G(\rho - 2\rho_{\Lambda}). &
\eeq

Here $\vv$ and $\rho$ are perturbed velocity and density. The dy\-na\-mi\-cal
effect of vacuum on the flow is taken into account in the Poisson equation by
the vacuum density $\rho_{\Lambda}$ which enters also the Eq.(\ref{eq:ddota}).
The velocity component $v_i$ depends on $x_i$ only.

Now we use the Lagrangian coordinates instead of the Eulerian coordinates in
accordance with the relations (prime denotes derivative on $\chi_1$)

\beq \label{eq:EulLag}
& \strut\disp \Dr {}{x_1}=\frac {1}{a+\delta'}
\Dr {}{\chi_1}, & \\
& \strut\disp \Dr {}{x_2}=\frac {1}{a}\Dr {}{\chi_2},\,\,
\Dr {}{x_3}=\frac {1}{a}\Dr {}{\chi_3}, & \\
& \strut\disp \Dr {}{t}\biggl|_{\vx}=\Dr {}{t}\biggl|_{\vchi}\!-
\frac {\dot {a}\chi_1+\dot {\delta}}{a+\delta'}\Dr {}{\chi_1}\!-\!
\frac {\dot {a}}{a}\left(\chi_2\Dr {}{\chi_2}+\chi_3\Dr {}{\chi_3}\right)\!. &
\eeq
As a result the continuity equation can be presented in the form
\begin{equation} \label{eq:conteq}
\frac {\dot {\rho}}{\rho} +\frac {\dot {a}+\dot {\delta}'}{a+\delta'} +
2\frac {\dot {a}}{a}=0.
\end{equation}
Its solution is
\begin{equation} \label{eq:contsol}
\rho =\frac {a_0^3}{a^2}\frac {\rho_0}{a+\delta'}.
\end{equation}

Similarly Eq.(\ref{eq:eqmove}) is reduced to the form
\begin{equation} \label{eq:moveeq}
\nabla\varphi=-\dot {\vv}.
\end{equation}
Substituting this relation into the Poisson equation gives
\begin{equation} \label{eq:releq}
-\frac {\ddot {a}}{a+\delta'}-2\frac {\ddot {a}}{a}= 4\pi G(\rho -
2\rho_{\Lambda}).
\end{equation}

With the use of Eqs.(\ref{eq:ddota}),(\ref{eq:dota2}) and (\ref{eq:contsol}),
we obtain from Eq.(\ref{eq:releq}) rather simple relation
\begin{equation} \label{eq:linrel}
\ddot {\delta'}=\frac {8\pi G}{3}\left(\rho_0\frac {a_0^3}{a^3}+
\rho_\Lambda\right)\delta'.
\end{equation}
Assuming, as usually, that this relation holds for both
$\disp\delta'$ and $\delta$, we cross then from the derivatives
with respect to $t$ to the derivatives with respect to
$s=a(t)/a_0$ and finally obtain the equation that controls the
time behaviour of the flow:
\beq \label{eq:eqpert}
& \strut\disp s^2\left(\rho_0+\rho_\Lambda s^3-\frac {kc^2}{a_0^2}
\frac {3}{8\pi G}s\right)\DR {^2\delta}{s^2} & \nonumber \\
& \strut\disp -s\left(\frac {1}{2}\rho_0-2\rho_\Lambda s^3\right)
\DR {\delta}{s}-(\rho_0+\rho_\Lambda s^3)\delta=0. &
\eeq

Eq.(\ref{eq:eqpert}) can be solved numerically with the use of a
standard procedure. It is most important that for the parabolic ($k =
0$) motion, an exact solution of the equation can be found in an
explicit analytical form.

\section {Exact solution}

Let us denote $q=\rho_\Lambda/\rho_0=\Omega_\Lambda/\Omega_0$ where
$\Omega_\Lambda$ and $\Omega_0$ are the present vacuum and matter
densities in the units of critical density. For $k=0$ one has
$\Omega_\Lambda+\Omega_0=1$ and
\begin{equation} \label{eq:Om0OmL1}
\Omega_\Lambda=\frac {q}{1+q},\,\,\Omega_0=\frac {1}{1+q}.
\end{equation}

Introducing $u =qs^3 = q [a(t)/a_0]^3$, where now
\begin{equation} \label{eq:afromt}
a(t)=a_0 q^{-1/3}u^{1/3}(t),\,\,u(t)=\sinh^2\left(\frac {3}{2}\alpha t\right),
\end{equation}
and $\alpha=\sqrt {\Omega_\Lambda}H_0=\sqrt {q/(1+q)}H_0=
\sqrt {8\pi G\rho_{\Lambda}/3}$, we transform Eq.(\ref{eq:eqpert}) to the form
\begin{equation} \label{eq:hypereq}
9u^2(1+u)\DR {^2\delta}{u^2}+9u\left(\frac {1}{2}+u\right)
\DR {\delta}{u}-(1+u)\delta=0,
\end{equation}

Eq.(\ref{eq:hypereq}) can be reduced to hypergeometric equation, and its
general solution is given in terms of hypergeometric functions. Accordingly,
one has the following solution:
\begin{equation} \label{eq:exacsol}
\delta = B(qs^3)f_\B(\chi_1) + C(qs^3)f_\C(\chi_1).
\end{equation}

The solution contains two modes of the perturbation flow (we call them 1 and
2) and includes $f_\B$ and $f_\C$ as arbitrary dimensionless functions of the
Lagrangian coordinate $\chi_1$. The second mode is given by an elementary
function:
\begin{equation} \label{eq:Cu}
C(u)= \frac {\sqrt {1+u}}{u^{1/6}}.
\end{equation}
As for the first mode, $B(u)$, it is given in terms of hypergeometric
functions:
\begin{equation} 
\label{eq:Bu}
B(u)=\left\{
\begin {array}{lcl}
\disp u^{2/3}F\left(1,\frac {1}{3},\frac {11}{6};-u\right) & {\rm if}
 & 0\leq u<1, \\
\disp \frac {u^{2/3}}{1+u}F\left(1,\frac {3}{2},\frac {11}{6};\frac {u}{1+u}
\right) & {\rm if} & u\sim 1, \\
\disp c_0 C(u)-\frac {5}{4}D(u) & {\rm if} & u>1, \\
\end {array}\right.
\end{equation}
where $\disp c_0=\frac{2}{\sqrt {\pi}}\Gamma\left(\frac{11}{6}\right)
\Gamma\left(\frac{2}{3}\right)$ and
\begin{equation} \label{eq:Du}
D(u)=\frac {1}{u^{1/3}}F\left(1,\frac {1}{6},\frac {5}{3};-\frac {1}{u}\right)=
\frac {4}{5}\left[c_0 C(u)-B(u)\right].
\end{equation}

Fig.1 demonstrates the behaviour of $B,\,C$ and $D$ as\, func-

\begin {figure}[ht]

{\small

\hspace {1mm} $\lg B,\,\lg C,\,\lg D $

{\psfig {file=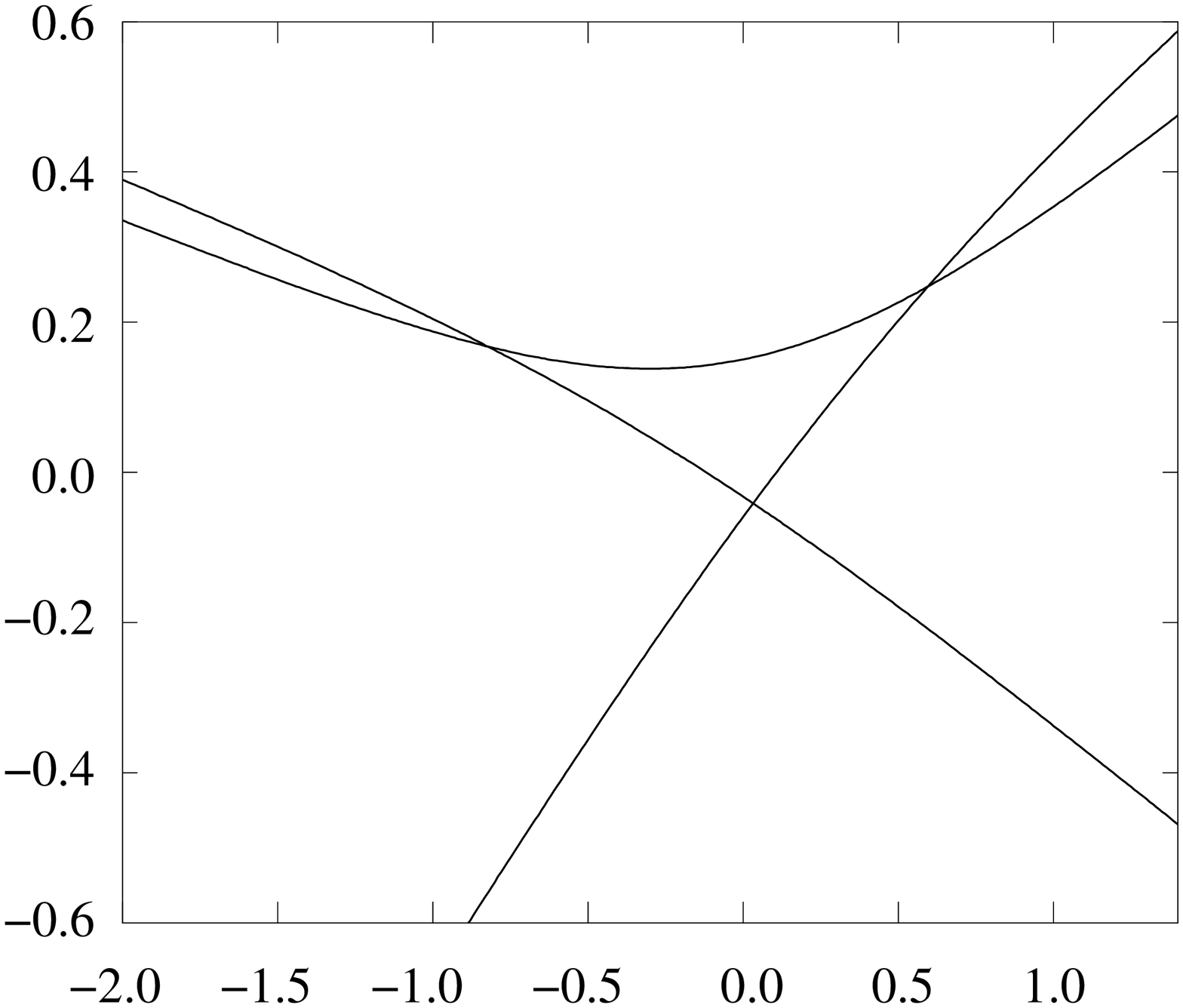,width=8cm}}

\vspace {-50mm}

\hspace {41mm} $C(u)$

\vspace {17mm}

\hspace {32mm} $B(u)$ \hspace {8mm} $D(u)$

\vspace {24mm}

\hspace {35mm} $\lg u$

}

\vspace {2mm}

\hspace {6mm} {\bf Fig.1.} The functions $B(u),\,C(u)$ and $D(u)$

\end {figure}

\noindent
tions of time variable $u$ (double logarithmic scale) in the most
interesting "transition" range of the argument $u$, e.g. in the era,
when the evolution of the flow transits from the initial epoch of
dynamical domination of matter gravity to the later epoch of
dynamical domination of vacuum anti-gravity.

It is also easy to see from Fig.1 the dependence of the functions on
the redshift $z$, since $u\!= q (\!a/a_0)^3\!=\! q (1+z)^{-3}$.

Simple asymptotic formulae for the three functions at the epochs before and
after the transition era are given in Table 1. In this table
the asymptotic dependence of $a$ and $u$ on time is also shown.

\bc
{\bf Table 1.} Asymptotic behaviour of $a(t),\,u(t),\,B(u),\,C(u),\,D(u)$

\medskip

\begin {tabular*}{8.0cm}{@{\extracolsep{\fill}}|c|c|c|}
\hline
 & $t\to 0$ & $t\to\infty$ \\
\hline
$a(t)$ & $a_0 q^{-1/3}(3\alpha t/2)^{2/3}$
 & $a_0q^{-1/3}2^{-2/3}e^{\alpha t}$ \\
$u(t)$    & $(3\alpha t/2)^2$   & $2^{-2}\exp(3\alpha t)$ \\
$B(u)$ & $u^{2/3}$         & $c_0 u^{1/3}$ \\
$C(u)$ & $u^{-1/6}$        & $u^{1/3}$ \\
$D(u)$ & $4c_0 u^{-1/6}/5$ & $u^{-1/3}$ \\
\hline
\end {tabular*}
\ec

\section {Discussion}

1. The exact solution (\ref{eq:exacsol}) contains the Zeldovich
solution of Eq.(1) and the exact solution of Eqs.(2,3) as
particular cases for $\rho_{\Lambda} = 0$. It is also seen from
Table 1 that Eqs.(1,2) represent an asymptotic of our solution in
the limit $t\to 0,\,a(t)\to 0$, when the dynamical effect of
vacuum can be neglected. Indeed, in this limit, the two modes of
our solution vary as
\begin{equation} \label{eq:BCas}
B (u(t)) \propto b (t) = t^{4/3}, \;\;
C (u(t)) \propto c (t) = t^{-1/3}.
\end{equation}

It may be easily found with this asymptotic that the arbitrary functions
of the solution of Eqs.(1,2) are related to the arbitrary
functions of our solution in a simple way:
\begin{equation} \label{eq:betgamf}
\beta (\chi_1)\!=\!\left(\frac {3}{2} \alpha\right)^{4/3} f_\B (\chi_1), \;\;
\gamma (\chi_1)\!=\!\left(\frac {3}{2}\alpha\right)^{-1/3} f_\C (\chi_1)\!.
\end{equation}

\smallskip

2. In the opposite asymptotic region, in the limit when $t\to\infty,\, a(t)
\to \infty$, vacuum dominates; since the gravity of matter
can be then neglected, matter may be considered as a gas of test
particles moving freely on the vacuum background. In this
limiting case, the time behaviour of the perturbation flow
differs drastically from the Zeldovich solution. Indeed, for
large times one has from our solution:
\begin{equation}
\delta \propto B(u) \propto C(u) \propto a(t)\propto e^{\alpha t}.
\end{equation}
As we see, the time behaviour of the perturbation flow
is exactly the same as that of the unperturbed Hubble flow, in
this limit. Accordingly, the whole nonlinear (and arbitrary
non-uniform in density --- see below) flow expands in a regular
isotropic manner with the Hubble linear law.

\smallskip

3. Our solution allows a special case, in which functions $f_\B$
and $f_\C$ are related to each other as $f_\B = - f_\C/c_0(=-4f_\D/5)$; the
asymptotic of this special solution at $a(t) \to \infty$
is different from both Eqs.(1,2) and the equation above:
\begin{equation}
\delta \propto D(u) \propto a(t)^{-1}\propto e^{-\alpha t}.
\end{equation}
The special solution describes the asymptotic "adiabatic cooling" of the
nonlinear flow.

\smallskip

4. According to Eq.(\ref{eq:contsol}), the perturbation of the
matter density relative to the perturbed density $ \disp\eps=\frac
{\delta\rho}{\rho}=-\frac{\delta'}{a}$. One may consider also the
density perturbation relative to the unperturbed density:
\begin{equation}
\eps_*=\frac {\delta\rho}{\rho_0}\frac {a_0^3}{a^3}=-
\frac {\delta'} {a+\delta'}=\frac {a}{a+\delta'}\eps.
\end{equation}

The asymptotics of the relative density perturbation $\eps_*$, when $t\to
0$ and $t\to\infty$, are given in Table 2 for the two modes of
the general solution (denoted in the Table as 1,2) and for the special
case 3.

\bc
{\bf Table 2.} Asymptotics of $\eps_*$

\medskip

\begin {tabular*}{9.0cm}{@{\extracolsep{\fill}}|c|c|c|}
\hline
Mode & $t\to 0$ & $t\to\infty$ \\
\hline
1 & $\disp-f_\B'\frac{\phantom {\Bigl|}q^{1/3}}{a_0\phantom {\Bigl|}}
\left(\frac{3}{2}\alpha t\right)^{2/3}$ & $\disp-f_\B'\frac{c_0q^{1/3}}
{a_0+f_\B'c_0q^{1/3}}$ \\
\hline
2 & $-1$ & $\disp-f_\C'\frac{\phantom {\Bigl|}q^{1/3}}
{\phantom {\Bigl|}a_0+f_\C'q^{1/3}}$ \\
\hline
3 & $-1$ & $\disp-f_\D'2^{4/3}\frac {\phantom {\Bigl|}q^{1/3}}
{\phantom {\Bigl|}a_0}e^{-2\alpha t}$ \\
\hline
\end {tabular*}
\ec

If one considers $\eps$ at small times, it may be seen that its behaviour is
the same as $\eps_*$ for the first mode of the solution. 
For the the second mode $\disp\eps\sim-f_\C'\frac{q^{1/3}}{a_0}\left(
\frac{3}{2}\alpha t\right)^{-1}$. For the third mode we must change $f_\C$ to
$f_\D$ and add factor $4c_0/5$.

When $t\to\infty$ then $\eps$ coincides with $\eps_*$ for mode 3 and does not
contain the second terms in denominators of modes 1, 2.

\smallskip

5. In the velocity perturbation, there is only one component,
$\disp\delta v_1=\dot {\delta}$, and the relative velocity perturbation
is
\begin{equation}
\delta_*v_1=\frac {\delta v_1}{\chi_1\dot {a}}=\frac {1}{\chi_1}
\Dr {\delta}{a}.
\end{equation}

The asymptotics of relative perturbations of velocity for the same
three cases 1,2,3, as in Table 2, are given in Table 3.

\bc
{\bf Table 3.} Asymptotics of $\delta_*v_1$

\medskip

\begin {tabular*}{9.2cm}{@{\extracolsep{\fill}}|c|c|c|}
\hline
Mode & $t\to 0$ & $t\to\infty$ \\
\hline
1 & $\disp\frac{f_\B\phantom {\bigl|}}{\chi_1}q^{1/3}
 \left(\frac{3}{2}\alpha t\right)^{2/3}$ &
$\disp\frac{f_\B}{\phantom {\bigl|}\chi_1}c_0q^{1/3}$ \\
\hline
2 & $\disp-\frac {f_\C\phantom {\bigl|}}{\chi_1}
\frac {q^{1/3}}{3}(\alpha t)^{-1}$
 & $\disp \frac {f_\C}{\phantom {\bigl|}\chi_1}q^{1/3}$ \\
\hline
3 & $\disp-\frac{4}{15}\frac {\phantom {\bigl|}f_\D}{\chi_1}c_0q^{1/3}
(\alpha t)^{-1}$ &
$\disp -\frac {f_\D}{\phantom {\bigl|}\chi_1}2^{4/3}q^{1/3}e^{-2\alpha t}$ \\
\hline
\end {tabular*}
\ec
(We omit the argument $\chi_1$ of functions $f$ in Table 3 and of
their derivatives in Table 2.)

This analysis shows that gravitational instability is terminated
and the nonlinear perturbations, that are growing at earlier
epoch, are then freezing out or even being "adiabatically cooled",
when vacuum starts to dominate dynamically; this effect is
clearly due to anti-gravity of vacuum.

\smallskip

6. It is worth to find a relation between the nonlinear solution
and the behaviour of linear perturbations on the vacuum background.
The general solution for linear perturbations may easily be found
with the method suggested by Zeldovich (1965) for the
Lifshits-type perturbations (see for comparison Heath 1977, Lahav
et al. 1991):
\begin{equation} \label{eq:solgen}
\delta=F_1(\chi_1) {\dot{a}(t)}\int \frac{d t}{\dot {a}^3(t)}  +
F_2(\chi_1) {\dot a(t)},
\end{equation}
where $F_1, F_2$ are arbitrary functions of $\chi_1$.
Two these linearly independent linear solutions are directly related to
modes 1,2 of the nonlinear solution obtained here if we use
$a(t)$ from (\ref{eq:afromt}):
\begin{equation}
\label{eq:CBa}
\dot {a}(t)\intl_0^t\frac {\d t}{\dot {a}^2(t)}\!=\!\frac {2}{5}\frac
{q^{1/3}} {a_0\alpha^2}B(u(t)),
\dot {a}(t)\!=\!a_0q^{-1/3}\alpha C(u(t)).
\end{equation}
For Mode 3 one has:
\begin{equation} \label{eq:Da}
D(u(t))=2\alpha^2a_0q^{-1/3}\dot {a}\intl_t^\infty\frac {\d t}{\dot {a}^2(t)}.
\end{equation}

\smallskip

7. Finally, on the basis of our solution,
a 3D `approximate' (in Zeldovich's (1970) sense) solution may be
suggested as a generalization of the Zeldovich ansatz by taking into
account the dynamical effect of cosmic vacuum:
\begin{equation}
\vx(\vchi,t) = a(t)\vchi + B(t) \vf_1(\vchi) + C(t) \vf_2(\vchi),
\end{equation}
where the time functions $B$ and $C$ are given by our nonlinear
solution, $\vf_1,\,\vf_2$ are arbitrary vector functions of the Lagrangian
3D coordinates $\vchi$.

We expect that this new generalized 3D approximation can
effectively be used for the theories of the large-scale structure
formation in $\Lambda CDM$ cosmologies, --  in the same way as
the Zeldovich original ansatz has been used for cosmologies with
zero cosmological constant.

\smallskip

It will be interesting to investigate the influence of radiation on
the evolution of perturbations.

\smallskip

{\it Acknowledgements}
This work was supported in part by the  Russian Leading Scientific  Schools
grant  00-15-96607 and Federal Program "Integration" project N 578.
A.C. appreciates a partial support from the Academy of Finland grant
"Galactic streams and dark matter structures".


\smallskip


\begin {thebibliography}{}

\bibitem[2000]{carrol} Carrol S.M., 2000, astro-ph/0004075.

\bibitem[1977]{heath} Heath D.J., 1977, MN RAS 179, 351.

\bibitem[1991]{lahav} Lahav O., Lilje P.B., Primack J.R., Rees M.J.,
1991, MN RAS 251, 128.

\bibitem[1946]{lif} Lifshits E.M., 1946, JETP 16, 587.

\bibitem[1993]{melott} Melott A.L., 1993, Comm. Astrophys. 17, 145.

\bibitem[1993]{peebles} Peebles P.J.E., 1993, Principles of Physical
Cosmology, Princeton University Press.

\bibitem[1999]{perl} Perlmutter S. et al., 1999. ApJ, 517, 565.

\bibitem[1998]{riess} Riess A., Filipenko A.V., Challis P., et al.,
1998, AJ 116, 1009.

\bibitem[1989]{shandzeld} Shandarin, S.F., Zel'dovich, Ya.B., 1989,
Rev. Mod. Phys. 61, 185.

\bibitem[2000]{Wang} Wang L. et al., 2000, Astrophys. J. 530, 17.

\bibitem[1977]{zch} Zentsova A.C., and Chernin A.D., 1980, Astrophysics
16, 108.

\bibitem[1965]{zeld65} Zeldovich Ya.B., 1965, Adv. Astron. Ap. 3, 241.

\bibitem[1970]{zeld70} Zeldovich Ya.B., 1970, A\&A 5, 84.

\end {thebibliography}
\end {document}